\documentclass[preprint2]{emulateapj}
\usepackage{subfigure}

\slugcomment{accepted by ApJ Letters}

\shorttitle{BRAVA-RR}
\shortauthors{Kunder et al.}

\begin{document}

\title{Before the Bar:  \\Kinematic Detection of A Spheroidal Metal-Poor Bulge Component}


\author{Andrea Kunder\altaffilmark{1},
R.~M.~Rich\altaffilmark{2},
A.~Koch\altaffilmark{3},
J.~Storm\altaffilmark{1},
D.~M.~Nataf\altaffilmark{4},
R.~De Propris\altaffilmark{5},
A.~R.~Walker\altaffilmark{6},
G.~Bono\altaffilmark{7,8},
C.~I.~Johnson\altaffilmark{9},
J.~Shen\altaffilmark{10},
Z.-Y.~Li\altaffilmark{10},
}
\altaffiltext{1}{Leibniz-Institut f\"{u}Ÿr Astrophysik Potsdam (AIP), An der Sternwarte 16, D-14482 Potsdam, Germany}
\altaffiltext{2}{Department of Physics and Astronomy, University of California at Los Angeles, Los Angeles, CA 90095-1562, USA}
\altaffiltext{3}{Physics Department, Lancaster University, Lancaster LA1 4YB, UK}
\altaffiltext{4}{Research School of Astronomy and Astrophysics, The Australian National University, Canberra, ACT 2611, Australia}
\altaffiltext{5}{Finnish Centre for Astronomy with ESO, University of Turku, 2150 Turku, Finland}
\altaffiltext{6}{Cerro Tololo Inter-American Observatory,  National Optical Astronomy Observatory, Casilla 603, La Serena, Chile}
\altaffiltext{7}{Dipartimento di Fisica, Universita di Roma Tor Vergata, vi a Della Ricerca Scientifica 1, 00133, Roma, Italy}
\altaffiltext{8}{INAF,  Rome  Astronomical  Observatory,  via  Frascati  33, 00040, Monte Porzio Catone, Italy}
\altaffiltext{9}{Harvard-Smithsonian Center for Astrophysics, Cambridge, MA 02138}
\altaffiltext{10}{Key Laboratory for Research in Galaxies and Cosmology, Shanghai Astronomical
Observatory, Chinese Academy of Sciences, 80 Nandan Road, Shanghai 200030, China}

\begin{abstract}
We present 947 radial velocities of RR Lyrae variable stars in four fields located toward the
Galactic bulge, observed within
the data from the ongoing Bulge RR Lyrae Radial Velocity Assay (BRAVA-RR).  
We show that these RR Lyrae stars exhibit hot
kinematics and null or negligible rotation and are therefore members of a separate population from 
the bar/pseudobulge that currently dominates the mass and luminosity of the inner Galaxy.  
Our  RR Lyrae stars predate these structures, and have metallicities, kinematics, and spatial 
distribution that are consistent with a ``classical" bulge, although we cannot yet completely 
rule out the possibility that they are the metal-poor tail of a more metal rich
($\rm [Fe/H]$$\sim$$-$1 dex)
halo-bulge population.  The complete catalog of radial velocities
for the BRAVA-RR stars is also published electronically.
\end{abstract}


\section{Introduction}
The majority of massive galaxies ($>$ 10$^{9}$$M_{\odot}$), similar to the Milky Way, have a
distinct rise in surface brightness above the disk, referred to as a bulge \citep{fisher11}.
Galaxy bulges are observed to either rotate rapidly like a disk, and are generally referred to as
pseudobulges, or they are dominated by random motions and are therefore 
pressure supported by a central velocity dispersion.  This latter type is referred to as a 
classical bulge \citep[e.g.,][]{kormendy82}.  That not all bulges are alike suggests that the 
bulge type of a galaxy carries significance for the galaxy's evolutionary history, such as 
its merger history and star formation efficiency \citep[e.g.,][]{martig12, obreja13, fiacconi15}.
The properties of the bulge in our Galaxy are therefore,
a fundamental parameter with which to understand the formation of the Milky Way.

The first wide-area spectroscopic surveys of the Milky Way bulge have shown that it
consists of a massive bar rotating as a solid body \citep{rich07, kunder12, ness13, zoccali14}.
The internal kinematics of these stars are consistent with at least 90\% of the inner Galaxy 
being part of a pseudobulge and lacking a pressure supported, classical-like 
bulge \citep{shen10, ness13}.
Recent studies have indicated a bimodal nature of bulges -- that two bulge populations, likened 
to classical and pseudo-bulges, can exist within a galaxy, with differences being in the 
relative proportions of the two \citep{obreja13, fisher16}.   There has accordingly been 
considerable debate about whether there is room for a
classical component in the bulge \citep{babusiaux10, zoccali08}.

The oldest and most metal poor stars (which may trace the dark matter) are thought 
to be found in the center of the Galaxy -- in the bulge but not sharing its kinematics 
and abundance patterns \citep{tumlinson10}.
Therefore, perhaps the greatest possibility of uncovering a classical component would 
be within the metal-poor bulge stars.  Unfortunately, spectroscopic surveys studying 
thousands of giants and red clump stars in the bulge have found that metal-poor stars 
in the bulge are rare, greatly limiting the use of metal-poor stars to uncover and probe 
a possible classical bulge component \citep{ness13, casey15, howes15, koch16}.  
Perhaps the easiest identifiable old, metal-poor bulge population are those horizontal 
branch stars that pulsate as RR Lyrae stars (RRLs).   
Since the absolute brightness of RRLs are known to within $\sim$10\%, the discovery 
of a significant population of RRLs toward the bulge permitted the first distance determination 
to the Galactic center from a stellar population \citep{baade46}.

In this paper, we report on the kinematics of a large sample of these stars in the Galactic bulge field.  
Our sample consists of RRLs with Galactocentric distances within $\sim$10\% 
of the distance of the Sun to the galactic center, so our sample
represents the typical ``bulge" RRL population.  

\section{Observations and Radial Velocity}
Observations were performed using the AAOmega multi-fiber spectrograph on 
the Anglo-Australian Telescope (AAT) on 
May 2013, June 2013, June 2014 and August 2015, in dual beam mode centered on 8600\AA, 
with the 580V and 1700D gratings to probe the Calcium Triplet
(NOAO PropID: 2014A-0143; PI: A. Kunder and NOAO PropID: 2015B-071; PI: A. Kunder).  
This covers the wavelength 
regime of about 8300\AA~to 8800\AA~at a resolution of R$\sim$10,000.  Exposure times 
were between one to two hours, and in general there are between 2 and 5 epochs for each RRL.
The 2013 observations were carried out in conjunction with a
bulge survey designed for detached red giant eclipsing binary twins (AAT: 2013A-05; PI: D. Nataf).
Extra fibers were available and resourcefully allocated to 95 bulge RRLs, and these RRLs 
have up to 15 epochs of observations.  

\begin{figure*}
\centering
\mbox{\subfigure{\includegraphics[height=8.4cm]{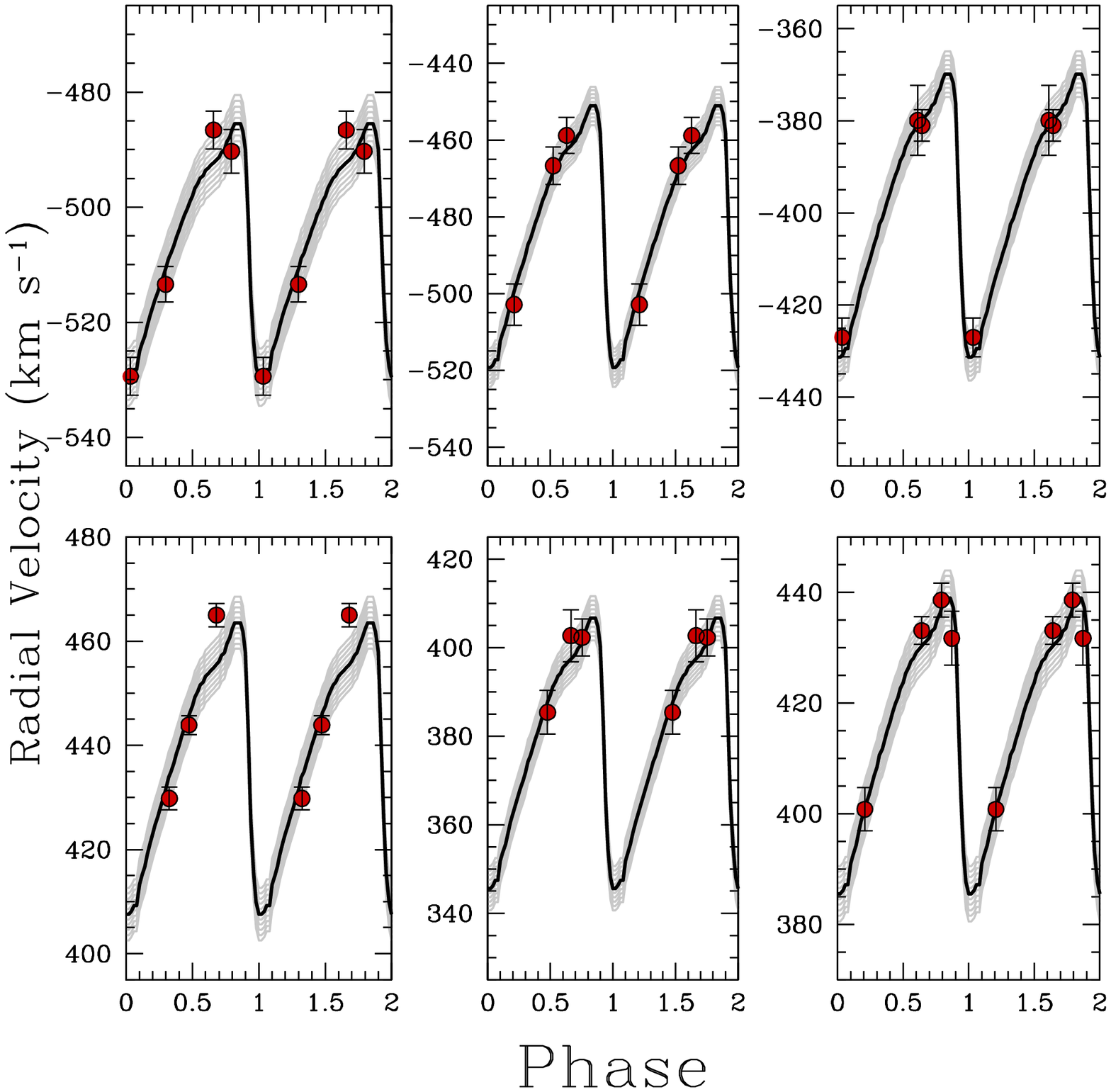}}\quad
\subfigure{\includegraphics[height=8.4cm]{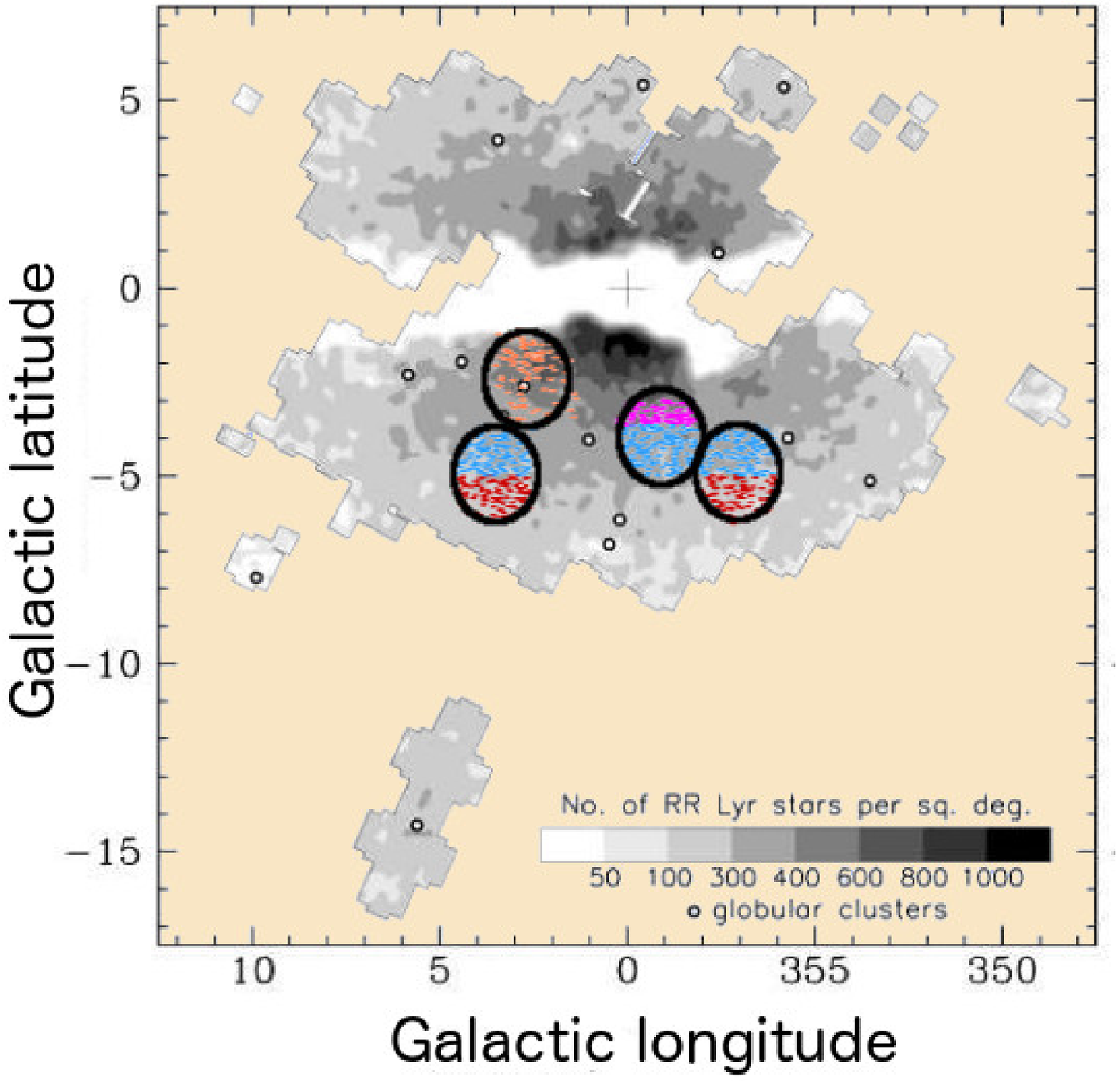} }
} 
\caption { {\it Left:} The line-of-sight radial velocity versus pulsational phase for a sample of BRAVA-RR observations; 
over-plotted is a fundamental mode RRL radial velocity template, scaled by its $V$-amplitude \citep[see
Equation~1, and also][]{liu91}.  The grey shaded area shows the 5~km~s$^{-1}$ uncertainty in the template,
which is also the typical uncertainty in our individual radial velocity measurements. 
{\it Right:} The spatial location of the OGLE RRLs in the Galactic bulge, with the 
RRLs presented here shown as bold symbols.  The observed RRLs are color coded to 
designate the strips of latitude they are separated into, with which to obtain their rotation 
curve (see Figure~2).  
}
\label{lcs}
\end{figure*}

The OGLE-III catalogue of RRLs \citep{pietrukowicz12} was used to select the targets.
We observed all bulge fundamental mode RRLs (RR0 Lyrae stars) 
that were free of a companion within a 2 arc second radius in our four fields.
These have been phased by the stars known period, and over-plotted with the 
radial velocity template from \citet{liu91} (Figure~\ref{lcs}).  This template is scaled 
using a correlation between the amplitudes of velocity curves and light curve:
\begin{equation}
A_{rv} = \frac{40.5\times V_{amp} + 42.7}{1.37}
\end{equation}
as shown in \citet{liu91}.
Because the $I$-amplitude of the OGLE stars is known much more precisely than the $V$-amplitude,
we use the relation $V_{amp}$ = $I_{amp}$ x 1.6 to find each stars 
$V$-amplitude \citep[see e.g., Table~3 in][]{kunder13}.  The so-called ``projection factor"
$p$= 1.37 is necessary because \citet{liu91} uses
pulsation velocities, which are related to observed radial velocities as
$v_{obs}=v_{puls}$/$p$.  The projection factor $p$ can range from 1.31 to 1.37 \citep{liu91, kovacs03},
and we adopted $p$= 1.37 as in \citet{sesar12}.

Due to our sampling techniques, almost all of our stars have at least two spectra at 
different phases, which facilitates the fitting of a radial velocity curve to
the measurements.  The zero-point in phase is fixed using the time of maximum 
brightness as reported by OGLE-IV \citep{soszynski14}
and the pulsation curve is shifted in radial velocity until it matches the observations.  
More weight is given to points that fall between $\phi =$0.0 - 0.6, as this is where
the uncertainty in the radial velocity shape of the template is minimised (see, e.g., Fig. 1 in Sesar 2012).
More weight is also given to the observations with a higher signal-to-noise, which 
also generally occurs between $\phi =$0.0 - 0.6.
The star's time-averaged velocity is determined by finding the velocity 
at $\phi_{obs}$=0.38 \citep{liu91}.  

We use 24 RRLs with well derived radial velocities to investigate how 2-3 epochs per
light curve affect our center-of-mass radial velocities.  The 24 RRLs are listed in
Table~\ref{rvstds}, and span a wide range of metallicities, pulsation periods and
amplitudes.  Because these 24 RRLs have individual radial velocity uncertainties that are
$\sim$1 km~s$^{-1}$, we first assign each radial velocity measurement a Gaussian uncertainty
distribution of 4~km~s$^{-1}$, to simulate the typical errors from the
observed BRAVA-RR observations.
We then use the \citet{liu91} template and decrease the number of epochs 
from $\sim$100 to 2 and measure how the center-of-mass radial velocity changes.
In general, the center-of-mass radial velocity changes by $\sim$2~km~s$^{-1}$
or less; only in the unusual cases where there were no epochs between $\phi =$0.0 - 0.6,
but instead the observations fell on the rising branch at phases greater than
0.85, did the center-of-mass radial velocity change by $\sim$5~km~s$^{-1}$.

Similar results have been shown previously;  for example, \citet{jeffery07} 
showed that center-of-mass radial velocities have a 
typical uncertainty of $\pm$1.5 km~s$^{-1}$ for variables observed at least 
three times when using the \citet{liu91} template.  
A visual inspection of the radial velocity curves of the 63 BRAVA-RR stars with 10 or 
more epochs of observations also indicates that the Liu (1991) template is 
sufficient to within 5 - 10 km$^{-1}$ to obtain a center-of-mass radial velocity.
Indeed, it is impressive how well the radial velocity template fits to the diverse
RRLs listed in Table~\ref{rvstds}.
We note that 87\% of our BRAVA-RR stars have 3 or more epochs of observations, 
making it statistically unlikely that the BRAVA-RR stars do not have observations
in the regime where the template most accurately aligns with the observed radial velocity measurements.
Therefore, our center-of-mass radial velocity 
uncertainties are 5-10 km~s$^{-1}$.  

Figure~\ref{lcs} shows example pulsation curves for the RRLs -- in particular, we show is
those with the most extreme radial velocities to illustrate this is a kinematically hot population.  
Table~\ref{lcpars} gives the OGLE-ID (1), the RA (2) and Dec (3) as provided by OGLE, 
the star's time-average velocity (4), the number of epochs used for the star's time-average velocity (5), 
the period of the star (6), the $V$-band magnitude (7),
the $I$-band magnitude (8) and the $I$-band amplitude (9) as calculated by OGLE, 
and lastly the photometric metallicity from the OGLE $I$-band light curve (10).

\section{Discussion}
\subsection{Rotation curve}
From spectroscopic observations of 947 RRLs in four 3 sq. deg. fields located toward the bulge
we plot the mean radial velocity and velocity dispersion for RRLs as a function 
of position (galactic latitude and longitude) in Figure~\ref{rotcurve}. These are found to be radically different 
from the trends traced by the more metal-rich red giants in the BRAVA and GIBS surveys \citep{kunder12, zoccali14}
as the RRLs show null rotation and hot (high velocity dispersion) kinematics. 
In the ARGOS survey one observes a slowly-rotating metal-poor population \citep{ness13}, 
which is hypothesized to arise from stellar contamination from disk and halo stars,
as it is only seen at high galactic latitude.
In contrast, our stars are at low galactic latitudes and their more certain distance 
estimates indicate they are within 1~kpc of the Galactic center, where the surface-density 
of bulge stars is usually dominant compared to the disk and halo.
We conclude that we are tracing an older, more spheroidal component in the inner Galaxy 
that may be likened to a classical bulge.  

\subsection{Metallicity distribution}
The $\rm [Fe/H]$ metallicity distribution in our sample
spans three orders of magnitude, with spectroscopic metallicities 
derived from the Calcium Triplet 8498 \AA~line \citep{wallerstein12}
ranging from $-$2.5 to super-solar metallicities, peaking at $\rm [Fe/H]$$\sim$$-$1~dex.
Therefore, as shown previously \citep{walker91}, the bulge field RRLs are
on average $\sim$1~dex more metal-poor than `normal' bar 
stars, yet some of the bulge field RRLs have metallicities that
overlap in abundance with the bar population.  
The bulge field RRLs are also more metal-rich than the stellar halo \citep[e.g.,][]{an13},
although the metallicity gradient observed in the field RRLs is consistent
with an inner bulge-halo at distances closer to the
Galactic center being more metal-rich \citep{suntzeff91}.

\begin{figure*}
\centering
\mbox{\subfigure{\includegraphics[height=8.4cm]{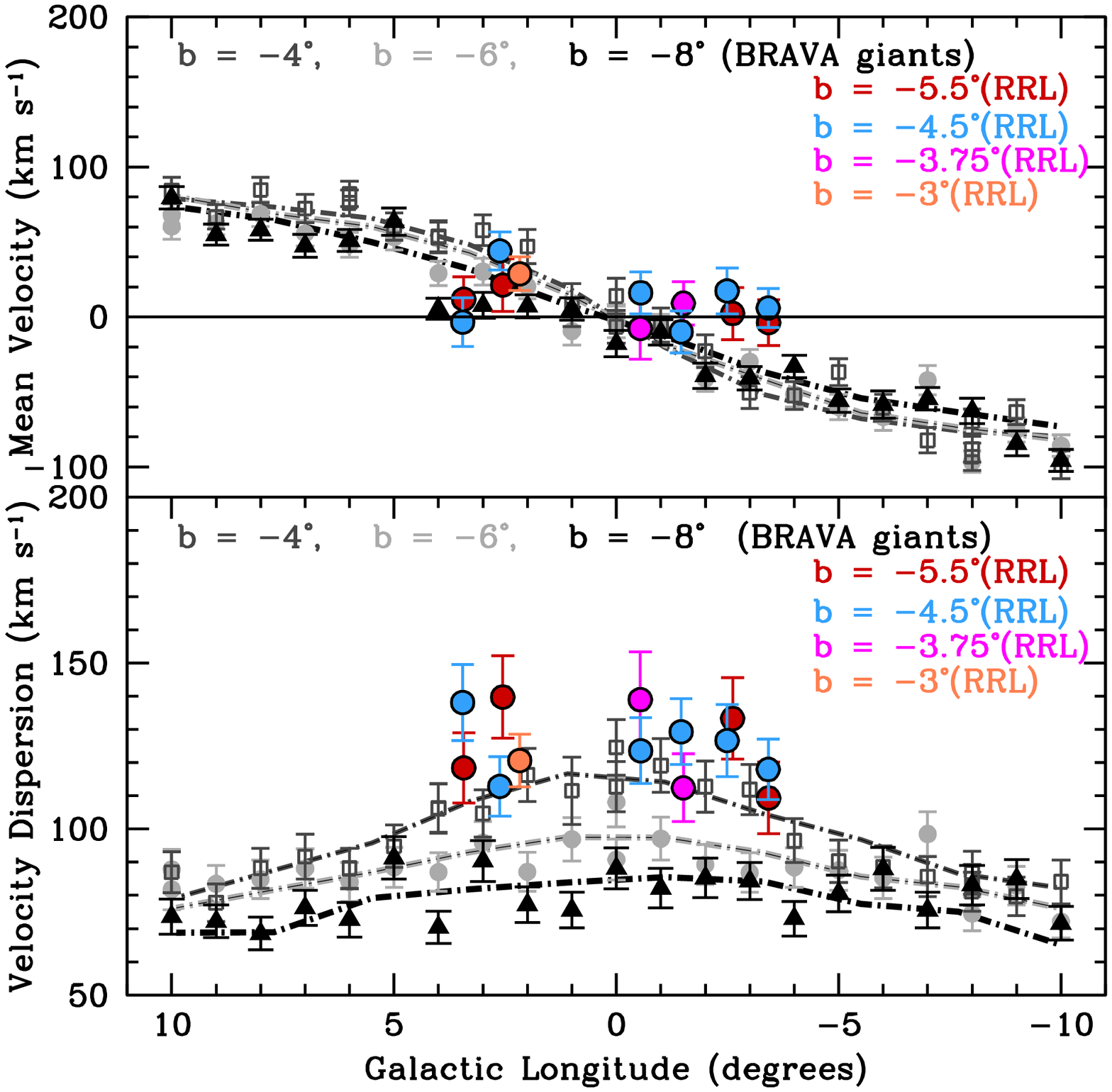}}\quad
\subfigure{\includegraphics[height=8.4cm]{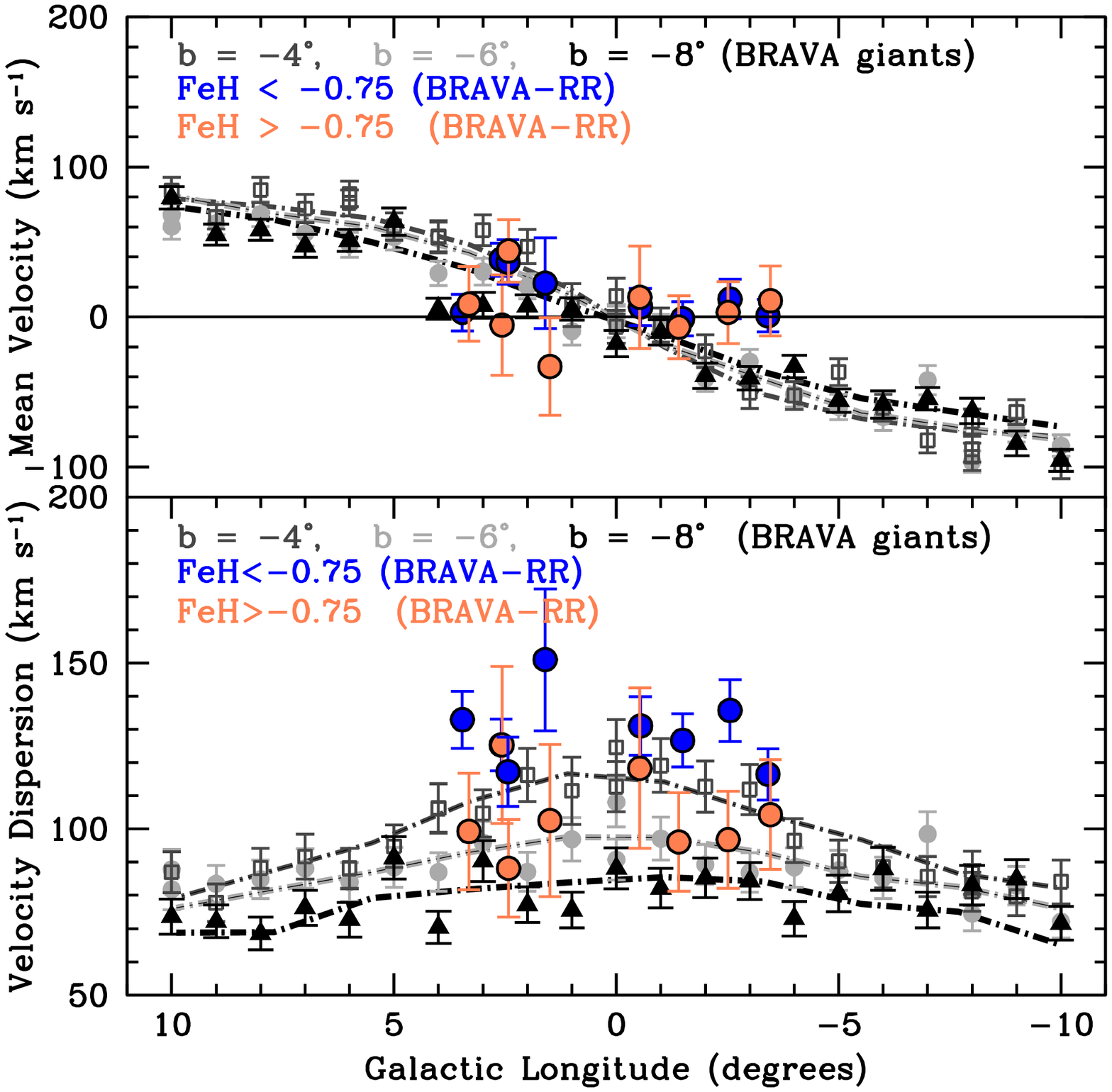} }
} 
\caption {
{\it Left:} The velocity dispersion profile (bottom) and rotation curve (top) for the RRLs we 
have already observed compared to that of the BRAVA giants at 
$b$ = $-$4$^\circ$, $-$6$^\circ$, and $-$8$^\circ$ strips \citep{kunder12}.
The bulge model showing these observations are consistent with a bulge
being formed from the disk is represented by the dashed lines \citep{shen10}.
The RRLs have kinematics clearly distinct from the bulge giants, and 
are a non-rotating population in the inner Galaxy. 
{\it Right:} The velocity dispersion profile (bottom) and rotation 
curve (top) for the RRLs seperated into metal-rich ($\rm [Fe/H]$$>$$-$0.75) 
and metal-poor  ($\rm [Fe/H]$$<$$-$0.75) stars.  The more metal-rich RRLs 
have metallicities that overlap with the pseudobulge red giants,
yet they still show no substantial rotation.  
\label{rotcurve}}
\end{figure*}

None of the bulge field RRL are extremely metal poor, in contrast
with what is predicted from a very old inner-halo \citep{tumlinson10, howes15}
there is presently no indication for many stars in our sample with 
$\rm [Fe/H]<-$3, although the evolutionary tracks for such metal poor stars 
may make it less likely for them to traverse through the instability strip,
thus becoming an RR Lyrae star \citep[e.g.,][see their Figure~1]{lee94}.  
The large metallicity spread is suggestive
of multiple populations within the bulge field RRL sample \citep{lee15}.
In Figure~\ref{rotcurve}, it is clear that the metal-rich RRLs have a smaller dispersion 
compared to the more metal-poor stars, which indicates there were likely
various RRL formation mechanisms in the bulge.  This
might plausibly be related to the two distinct bulge field RRL sequences in the 
period-amplitude diagram \citep{pietrukowicz15}
as well as the suggestion that only in the most central part (inner 1 kpc)
of the bulge do the RRLs exhibit a weak bar-like substructure \citep{pietrukowicz12, dekany13}.

We note that the BRAVA-RR spectroscopic metallicities are still being finalized, and this will
be the topic of a subsequent BRAVA-RR paper.  
Therefore throughout this paper we use photometric metallicities obtained from a
linear metallicity relationship in the pulsational period and phase difference between
the first and third harmonic $\phi_{31}$ (where $\phi_{nm} = m\phi_{n} - n\phi_{m}$)
in a Fourier decomposition of the OGLE $I$-band
lightcurves \citep{smolec05}.  These photometric metallicites are placed on the \citet{carretta09} 
metallicity scale.
The plots, however, does not change significantly when using our
preliminary CaT abundances.

\subsection{Mass estimate}
We can estimate the mass of the 'old' inner Galaxy component by comparing the relative 
numbers of red clump giants (metal rich horizontal branch stars) and RRLs in 
the OGLE-III survey \citep{pietrukowicz12, nataf13}. Here we assume that all the red clump stars are part of the 
rotating bar, whereas all the RRL are part of a non-rotating component (e.g., a classical bulge).  
We also presume that RRLs have the same lifetime as red clump stars and were 
formed from stellar populations with the same IMF as the red clump stars.
The kinematically hot component we recover in this 
paper then amounts to $\sim$1\% of the total central mass.
A similar mass is calculated from 
the fuel consumption theorem \citep{renzini86}, assuming again all the RRL 
are part of a non-rotating component and have a narrow range in age.  
This is broadly consistent with current bulge formation models, 
which predict that no more than $\sim$5\% of a merger-generated bulge,
which is slowly rotating and dispersion-supported, may 
exist within the Milky Way bulge \citep{shen10, ness13, dimatteo15}.
Although such a small mass may pose a challenge in understanding how this central component 
could remain stationary in the much more massive bar potential \citep{saha13}, some dynamical studies 
do suggest that a hot population
is only weakly affected, if at all, by the bar dynamics \citep{minchev12}. 

\subsection{Interpretation}
Our velocities rule out that possibility that the majority of RRL in the direction
of the bulge are part of the bar.  Given the ages of RRLs \citep{walker89, lee92}, 
this indicates that the 
inner Galaxy component traced out by the BRAVA-RR stars is at least $\sim$1~Gyr 
older than the dominant bar population.

\begin{figure*}
\centering
\mbox{\subfigure{\includegraphics[height=8.4cm]{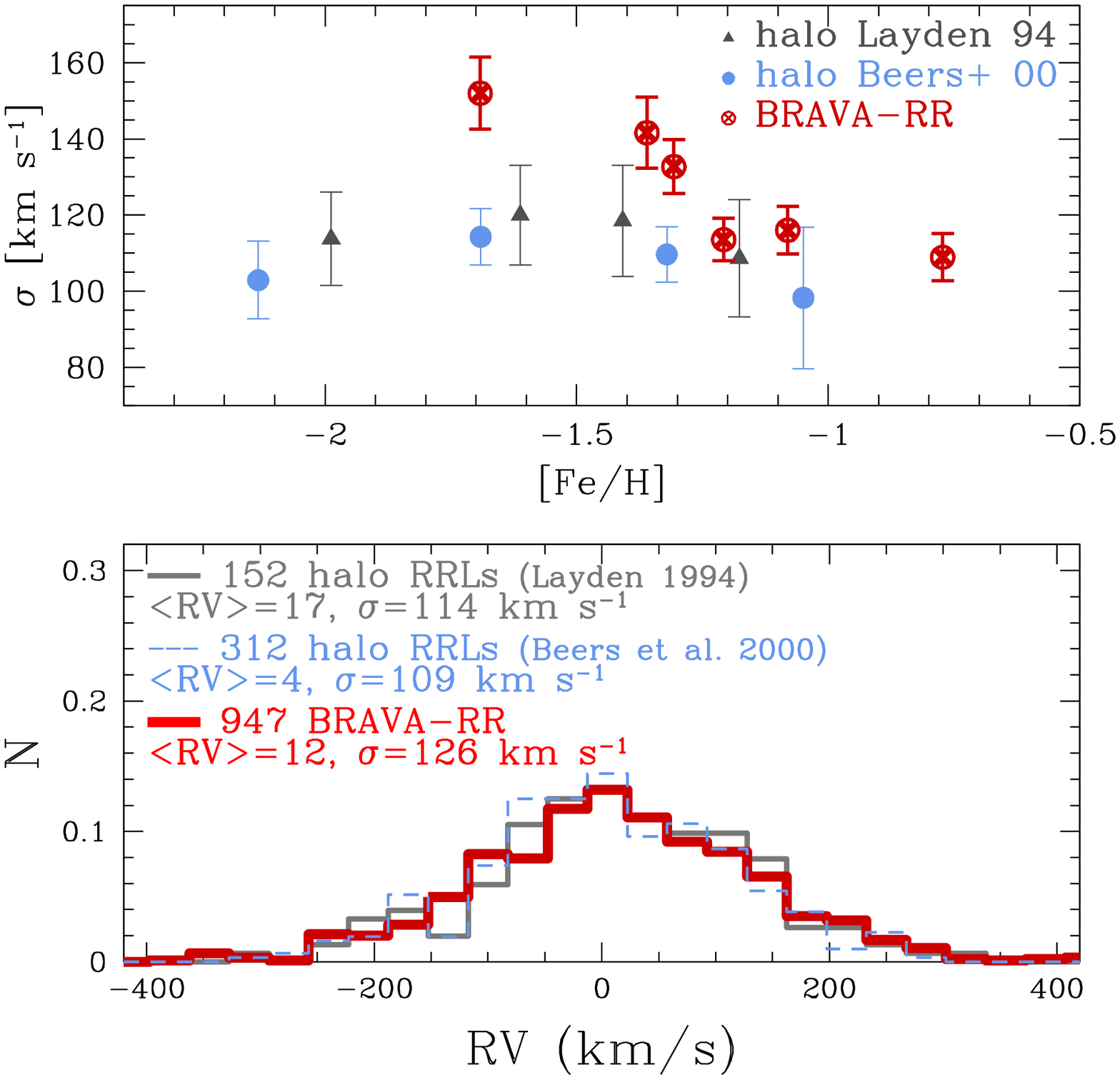}}\quad
\subfigure{\includegraphics[height=8.4cm]{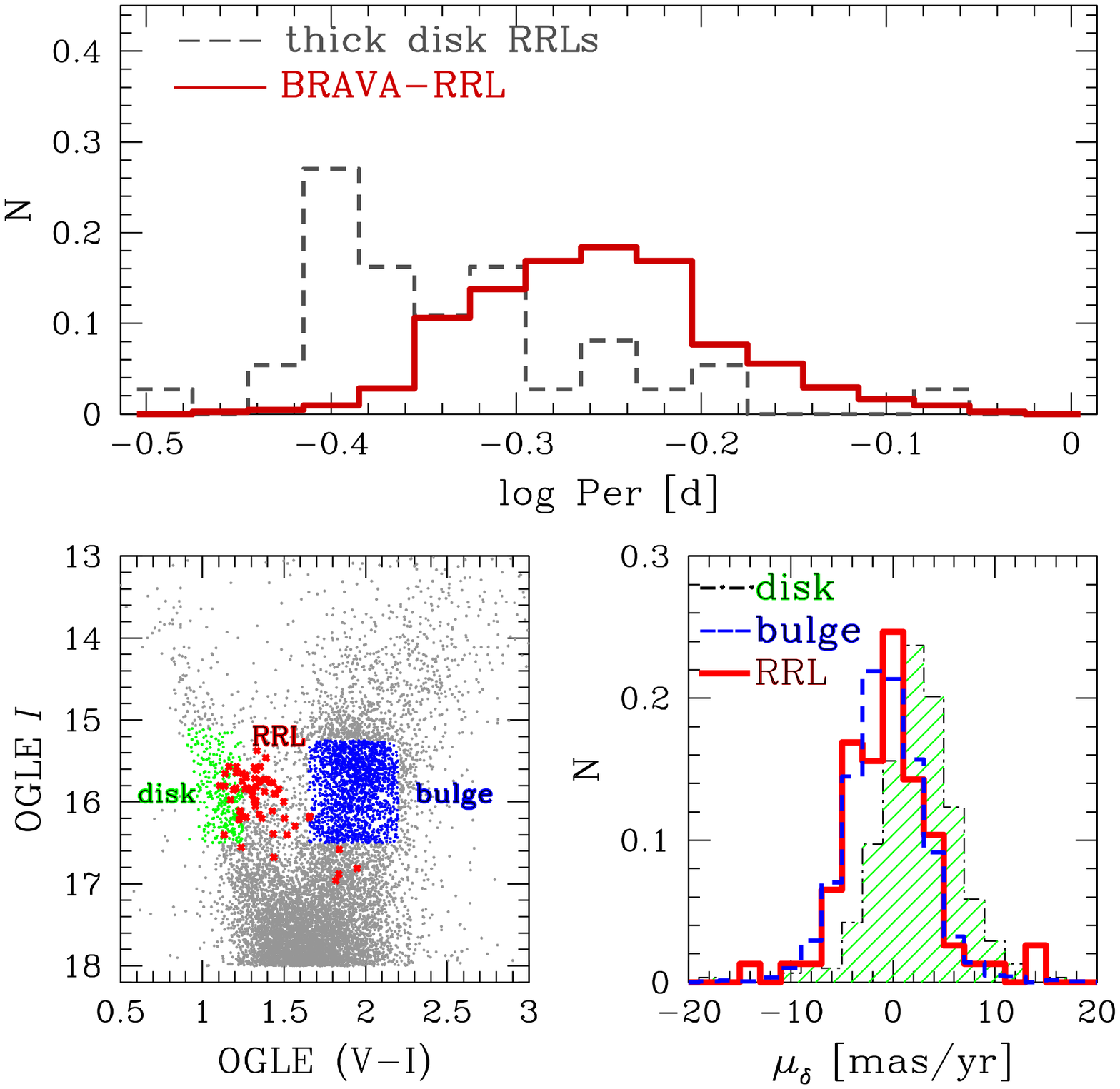} }
} 
\caption {
{\it Left top:} The velocity distribution of the BRAVA-RR stars and the halo RRLs in the \citet{layden94}
and \citet{beers00} sample as a function of $\rm [Fe/H]$.
{\it Left bottom:} The radial velocity distributions showing the kinematically selected halo RRL stars 
from \citet{layden94}, the non-kinematically selected metal-poor ($\rm [Fe/H]$ $<$ $-$1~dex) RRL 
sample from \citet{beers00}, and the BRAVA-RR stars presented here. \\
{\it Right top:} The period distribution of local thick disk RRLs selected kinematically by \citet{layden94, layden95} compared
to the period distribution of the BRAVA-RR stars.  The bulge RRLs have longer periods as compared to the
thick disk RRLs.
{\it Bottom Left:} The CMD of a typical bulge field in the OGLE-II catalog showing the
separation of disk (green) and bulge (blue) populations; the magnitudes and colors of the 77 BRAVA-RR
stars with OGLE-II proper motions are shown as red crosses.
{\it Bottom Right:}  The histogram of the OGLE-II proper motions for the disk (green), bulge (blue), and BRAVA-RR stars (red)
(see left panel).  The BRAVA-RR stars follow the proper motion distribution expected for a typical
bulge population. 
\label{diskcomp}}
\end{figure*}

It may be that the RRL stars toward the bulge are actually an inner 
halo-bulge sample, as originally speculated in the early 1990s \citep[e.g.,][]{minniti94}
and as at least one RRL orbit toward the Galactic bulge seems to 
indicate \citep{kunder15}.  However, the velocity dispersion of the bulge RRLs 
is $\sim$10~km~s$^{-1}$ larger than that seen in both the local RRL halo sample 
\citep{layden94, beers00} and from other halo star samples \citep[e.g.][]{battaglia06, brown09}.
In fact, we are not aware of any other stellar population in the galaxy with a larger 
velocity dispersion than that of the BRAVA-RR stars.
The decrease in velocity dispersion with metallicity (seen in Figure~2 and Figure~3) is also 
characteristic of stars located in the bulge regions \citep[e.g.,][]{rich90, johnson11, babusiaux10}.  
In contrast, the velocity dispersion of halo stars either does not change \citep[e.g.,][]{norris86},
or does not change as significantly \citep[e.g.,][]{chiba00, kafle13} as a function of metal-abundance.
A comparison of velocity dispersion with $\rm [Fe/H]$ for the BRAVA-RR stars and
for the halo RRL star sample of \citet{layden94} and
\citet{beers00} is shown explicitly in Figure~\ref{diskcomp} (left panel).
If the RRL stars toward the bulge are an inner halo-bulge population, this 
component would be the most metal-rich halo population
identified in the Galaxy, with a mean $\rm [Fe/H]$$\sim$$-1$~dex, compared to the
inner halo ($\rm [Fe/H]$$\sim$$-1.6$~dex)
and the outer halo  ($\rm [Fe/H]$$\sim$$-2.2$~dex) \citep[e.g.,][]{an13}.

Recent studies have indicated that a kinematically warmer component
associated with the Galactic thick disk could be present in the bulge, and 
would not be part of the bar structure traced out by the majority of the bulge 
red giants \citep[e.g.,][]{dimatteo15}.  We therefore verified if the bulge RRLs 
have properties that could link them to the thick disk.  However, as seen in 
Figure~\ref{diskcomp} (left panel), OGLE-II proper motions \citep{sumi04}
of our observed RRLs set them apart from that of the disk,
not surprising, as it is known that only $\sim$20\% of RRLs in the Milky Way 
reside in the (thick) disk \citep{layden95}.  Figure~\ref{diskcomp} also illustrates 
how the period distribution of the bulge field RRLs is shifted to longer periods in 
comparison to the RRL kinematically identified by \citet{layden94, layden95}
as belonging to the thick disk.
Although the local thick disk sample is small (37 stars total), from an RRL 
pulsational tagging stand point it appears unlikely that the BRAVA-RR component 
was formed in a similar manner to that of the Milky Way thick disk.

\section{Conclusions}
It has proven extremely difficult to disentangle the formation history of the inner Galaxy.  
RRLs are the only luminous evolved stars for which it is possible to place a time stamp:
these stars are older than 11 Gyr \citep{walker89}.
It appears that in the RRL population toward the Galactic bulge, we can observe a
distinct stage of the formation of the inner Galaxy that was antecedent to the formation of the bar. 
This is in agreement with an axisymmetric geometry described using
near-infrared VVV observations \citep{dekany13}, and is in contrast to the view provided from optical OGLE
photometry in which the RRLs appear to follow the elongated spatial distribution of the bar \citep{pietrukowicz15}.

The different kinematics of the ``bulge" field RRLs and the majority of the bulge giants supports the 
claim that galaxies may harbor two populations in the inner Galaxy, which may be 
likened to classical and pseudobulges, with differences being with the fraction of the two \citep{obreja13}.
Within the RRLs population toward the direction of the bulge, we can probe an early epoch inner Galaxy that 
was formed before the massive disk secularly evolved into the bar.
Detailed models of the halo, thick disk and bulge components $\sim$1~kpc from the Galactic center, 
an understanding of the elemental abundances of the RRLs, as well as a large sample of accurate proper motions 
for our BRAVA-RR stars, will
help distinguish if the ``bulge" RRLs reside in a classical-like bulge, or are part of a
different Milky Way component, such as a metal-rich inner halo-bulge.

\acknowledgements
We thank the Australian Astronomical Observatory, which have made these observations possible. 
This research was supported in part by the National Science Foundation under Grant No. NSF PHY11-25915.
This work was supported by Sonderforschungsbereich SFB 881 "The Milky 
Way System" (subprojects A4, A5, A8) of the German Research Foundation (DFG).
RMR acknowledges support from grant
AST-1413755  from the National Science Foundation.
C.I.J. gratefully acknowledges support from the Clay Fellowship, administered by the Smithsonian Astrophysical Observatory.

\clearpage

\begin{table}
\centering
\caption{Radial velocity light curves of local RR Lyrae stars}
\label{rvstds}
\begin{tabular}{p{0.5in}p{0.75in}p{0.38in}p{1.95in}} \\ \hline
Name & log Period (d) & $\rm [Fe/H]$ & Source \\ 
\hline
WY~Ant & -0.240838 & -1.25 & \citet{skillen93} \\
X~Ari & -0.186314 & -2.20 & \citet{jones87} \\
RR~Cet & -0.257245 & -1.25 & \citet{liujanes89} \\
UU~Cet & -0.217469 & -1.00 & \citet{clementini90} \\
W~Crt & -0.383976 & -0.70 & \citet{skillen93} \\
DX~Del & -0.325491 & -0.20 & \citet{meylan86} \\
RX~Eri & -0.231180 & -1.40 & \citet{liujanes89} \\
SS~Leo & -0.203186 & -1.51 & \citet{fernley90} \\
RV~Oct & -0.243239 & -1.75 & \citet{skillen93} \\
V445~Oph & -0.401187 & -0.39 & \citet{fernley90} \\
AV~Peg & -0.408518 & 0.00 & \citet{liujanes89} \\
RV~Phe & -0.224454 & -1.50 & \citet{cacciari87, cacciari89} \\
BB~Pup & -0.318267 & -0.60 & \citet{skillen93} \\
VY~Ser & -0.146245 & -1.80 & \citet{carney84} \\
W~Tuc & -0.192305 & -1.35 & \citet{clementini90} \\
UU~Vir & -0.322752 & -0.55 & \citet{jones88, liujanes89} \\
47Tuc-V9 & -0.132620 & -0.71 & \citet{storm94} \\
M4-V2 & -0.271092 & -1.30 & \citet{liujanes90} \\
M4-V32 & -0.237240 & -1.30 & \citet{liujanes90} \\
M4-V33 & -0.211245 & -1.30 & \citet{liujanes90} \\
M5-V8 & -0.262617 & -1.40 & \citet{storm92} \\
M5-V28 & -0.264455 & -1.40 & \citet{storm92} \\
M92-V1 & -0.153171 & -2.24 & \citet{storm92} \\
M92-V3 & -0.195575 & -2.24 & \citet{storm92} \\
\hline
\end{tabular}
\end{table}

\begin{table}
\begin{scriptsize}
\centering
\caption{Radial velocities of BRAVA-RR stars}
\label{lcpars}
\begin{tabular}{p{0.55in}p{0.55in}p{0.58in}p{0.55in}p{0.55in}p{0.58in}p{0.28in}p{0.28in}p{0.28in}p{0.28in}} \\ \hline
OGLE ID & R.A. (J2000.0) & Decl. (J2000.0) & $\rm HRV_{\phi=0.38}$ (km~s$^{-1}$) & \# Epochs & Period (d) & $(V)_{mag}$ & $(I)_{mag}$ & $I_{amp}$ & $\rm [Fe/H]_{phot}$ \\ 
06032 & 17 53 15.15 & -34 10 21.3 &  13 & 3 & 0.53840530 & 17.442 & 16.209 & 0.67 & -0.98 \\
06138 & 17 53 23.86 & -34 10 48.5 & -142 & 3 & 0.51685809 & 16.929 & 15.763 & 0.49 & -0.59 \\
06166 & 17 53 26.77 & -34  08 58.1 &  -3 & 3 & 0.53398387 & 16.986 & 15.637 & 0.62 & -1.00 \\
06171 & 17 53 27.14 & -33 57 53.7 &  59 & 3 & 0.60629585 & 17.541 & 16.006 & 0.27 & -0.76 \\
06197 & 17 53 28.89 & -34 18 09.3 &  27 & 4 & 0.48368631 & 16.791 & 15.696 & 0.70 & -1.08 \\
06227 & 17 53 30.50 & -34 24 57.0 & -12 & 6 & 0.50614560 & 16.841 & 15.760 & 0.59 & -1.15 \\
06257 & 17 53 33.61 & -33 55 18.0 & 130 & 3 & 0.55305696 & 17.951 & 16.444 & 0.56 & -0.91 \\
06280 & 17 53 35.52 & -34 05 14.5 &  17 & 2 & 0.50637949 & 18.224 & 16.740 & 0.17 & 0.17 \\
06377 & 17 53 43.29 & -34 07 09.9 & -191 & 3 & 0.54579032 & 17.606 & 16.186 & 0.56 & -0.91 \\
06382 & 17 53 43.89 & -34 12 24.0 & -219 & 4 & 0.68896691 & 16.840 & 15.674 & 0.63 & -1.32 \\
\hline
\end{tabular}
\end{scriptsize}
\end{table}
\clearpage

\acknowledgments

\clearpage

\end{document}